\documentstyle[aps,psfig,multicol]{revtex}
\newcommand{\bcols}{\ifpreprintsty\else\begin{multicols}{2}\fi}
\newcommand{\ecols}{\ifpreprintsty\else\end{multicols}\fi}
\begin{document}
\draft
\title{Thermal radiation and amplified spontaneous emission from a random
medium}
\author{C. W. J. Beenakker}
\address{Instituut-Lorentz, Leiden University,
P.O. Box 9506, 2300 RA Leiden, The Netherlands}
\date{January 1998}
\maketitle
\begin{abstract}
We compute the statistics of thermal emission from systems in which the
radiation is scattered chaotically, by relating the photo\-count distribution
to the scattering matrix --- whose statistical properties are known from
random-matrix theory. We find that the super-Poissonian noise is that of a
black body with a reduced number of degrees of freedom. The general theory is
applied to a disordered slab and to a chaotic cavity, and is extended to
include amplifying as well as absorbing systems. We predict an excess noise of
amplified spontaneous emission in a random laser below the laser threshold.
\end{abstract}
\pacs{PACS numbers: 42.50.Ar, 05.45.+b, 42.25.Bs, 78.45.+h}
\bcols

The emission of photons by matter in thermal equilibrium is not a series of
independent events. The textbook example is black-body radiation
\cite{Lou83,Man95}: Consider a system in thermal equilibrium (temperature $T$)
that fully absorbs any incident radiation in $N(\omega)$ propagating modes
within a frequency interval $\delta\omega$ around $\omega$. A photo\-detector
counts the emission of $n$ photons in this frequency interval during a long
time $t\gg 1/\delta\omega$. The probability distribution $P(n)$ is given by the
negative-binomial distribution with $\nu=Nt\delta\omega/2\pi$ degrees of
freedom,
\begin{equation}
P(n)\propto{n+\nu-1\choose n}\exp(-n\hbar\omega/k_{\rm B}T).\label{binomial}
\end{equation}
The binomial coefficient counts the number of partitions of $n$ bosons among
$\nu$ states. The mean photo\-count $\bar{n}=\nu f$ is proportional to the
Bose-Einstein function
\begin{equation}
f(\omega,T)=[\exp(\hbar\omega/k_{\rm B}T)-1]^{-1}.\label{BEfunction}
\end{equation}
In the limit $\bar{n}/\nu\rightarrow 0$, Eq.\ (\ref{binomial}) approaches the
Poisson distribution $P(n)\propto\bar{n}^{n}/n!$ of independent photo\-counts.
The Poisson distribution has variance ${\rm Var}\,n=\bar{n}$ equal to its mean.
The negative-binomial distribution describes photo\-counts that occur in
``bunches'', leading to an increase of the variance by a factor
$1+\bar{n}/\nu$. These basic facts are known since the beginning of this
century \cite{Ein09}.

Thermal radiation is also referred to as ``chaotic radiation''
\cite{Lou83,Man95}. In recent years the word ``chaotic'' has entered optics in
a different context, to describe systems that scatter radiation in an
irregular, random way \cite{Noc97}. Such systems, typically, have weak
absorption, so they are far from being black bodies. Two recent papers have
studied deviations from black-body radiation in the case of one-dimensional
scattering \cite{Gru96,Art97}, but chaotic systems are intrinsically not
one-dimensional. What, then, is the statistics of the chaotic radiation
resulting from chaotic scattering? That is the problem addressed in this paper.

This problem is significant for more than one reason. First of all, thermal
emission is a fundamental property of a system. Deviations from the black-body
limit contain information on chaotic scattering that can not be obtained from
classical scattering experiments. Most studies of the optical properties of
random media have been restricted to classical optics \cite{She90}. The
similarity between the classical wave equation and the Schr\"{o}dinger equation
has permitted the transfer to classical optics of powerful theoretical
techniques from condensed matter physics \cite{Efe97}. Our solution of the
thermal-radiation problem demonstrates how one of these techniques, the method
of random-matrix theory \cite{Meh91}, can be applied to quantum optics. That is
the second reason for the significance of this problem. The third reason is the
recent interest in amplifying random media, motivated by possible applications
as a ``random laser'' \cite{Law94,Wie97}. A linear amplifier can be thought of
as being in thermal equilibrium at a negative temperature \cite{Jef93}, so that
our theory of thermal radiation can also deal with amplified spontaneous
emission.

\begin{figure}
\hspace*{\fill}
\psfig{figure=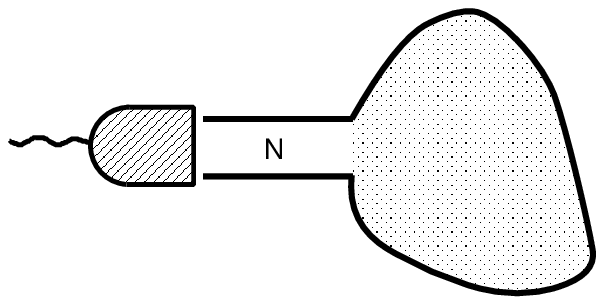,width=8cm}
\hspace*{\fill}
\smallskip\\
\refstepcounter{figure}
\label{diagram}
Fig.\ \ref{diagram} ---
Schematic diagram of a random medium (dotted) connected to a photo\-detector
(shaded) via an $N$-mode wave\-guide.
\end{figure}

We start with the formulation and solution of the problem in general form, and
then turn to specific applications. We consider a random medium coupled to a
photo\-detector via a wave\-guide (in vacuum) with $N(\omega)$ propagating
modes (counting polarizations) at frequency $\omega$. (See Fig.\
\ref{diagram}.) We assume that any Brownian motion of the scattering centra in
the random medium can be disregarded on the time scale of the measurements. The
scattering rate is denoted by $1/\tau_{\rm s}$, and the absorption or
amplification rate by $1/\tau_{\rm a}$. To quantize the electromagnetic field
we use the method of input--output relations developed by Gruner and Welsch
\cite{Gru96} and by Loudon and coworkers \cite{Art97,Jef93,Mat95}. The incoming
and outgoing modes in the wave\-guide are represented by two $N$-component
vectors of annihilation operators $a^{\rm in}(\omega)$, $a^{\rm out}(\omega)$.
They satisfy the commutation relations
\begin{equation}
[a_{n}^{\vphantom{\dagger}}(\omega),a_{m}^{\dagger}(\omega')]=
\delta_{nm}\delta(\omega-\omega'),\;\;
[a_{n}(\omega),a_{m}(\omega')]=0,\label{cra}
\end{equation}
for $a=a^{\rm in}$ or $a=a^{\rm out}$. The input-output relations take the form
\cite{Gru96,Art97,Jef93,Mat95}
\begin{equation}
a^{\rm out}=S\cdot a^{\rm in}+U\cdot b+V\cdot c^{\dagger},\label{aoutSain}
\end{equation}
with $S(\omega)$ the $N\times N$ scattering matrix. The boson operators $b$ and
$c$ satisfy Eq.\ (\ref{cra}) provided
\begin{equation}
U\cdot U^{\dagger}-V\cdot V^{\dagger}=\openone-S\cdot
S^{\dagger}\label{UVrelation}
\end{equation}
($\openone$ denoting the $N\times N$ unit matrix). The matrix $\openone-S\cdot
S^{\dagger}$ is positive definite in an absorbing medium, so we can put $V=0$.
Conversely, in an amplifying medium $\openone-S\cdot S^{\dagger}$ is negative
definite, so we can put $U=0$. This determines $U,V$ up to a unitary
transformation. All our final expressions depend only on the combination
$U\cdot U^{\dagger}-V\cdot V^{\dagger}$, so that any freedom in the choice of
$U,V$ is irrelevant once the scattering matrix is fixed.

Eq.\ (\ref{UVrelation}) can be understood as a fluctuation-dissipation
relation: The left-hand side accounts for quantum fluctuations in the
electromagnetic field due to spontaneous emission or absorption of photons, the
right-hand side accounts for dissipation due to absorption (or stimulated
emission in the case of an amplifying medium). Eq.\ (\ref{UVrelation}) also
represents a link between classical optics (the scattering matrix $S$) and
quantum optics (the quantum fluctuation matrices $U,V$).

In an absorbing medium, the operator $b$ accounts for thermal emission with
expectation value
\begin{equation}
\langle b_{n}^{\dagger}(\omega)b_{m}^{\vphantom{\dagger}}(\omega')\rangle=
\delta_{nm}\delta(\omega-\omega')f(\omega,T).\label{meanb}
\end{equation}
The inverted oscillator $c$ accounts for spontaneous emission in an amplifying
medium. We consider the regime of linear amplification, below the laser
threshold. Formally, this regime can be described by a thermal distribution at
negative temperature $-T$,
\begin{equation}
\langle c_{n}^{\vphantom{\dagger}}(\omega)c_{m}^{\dagger}(\omega')\rangle=
-\delta_{nm}\delta(\omega-\omega')f(\omega,-T),\label{meanc}
\end{equation}
the zero-temperature limit corresponding to a complete population inversion
\cite{Jef93}.
Higher order expectation values are obtained by pairwise averaging, as one
would do for Gaussian variables, after having brought the operators into normal
order.

The incoming radiation is in the vacuum state, while the outgoing radiation is
collected by a photo\-detector \cite{alpha}. The probability that $n$ photons
are counted in a time $t$ is given by \cite{Gla63,Kel64}
\begin{eqnarray}
&&P(n)=\frac{1}{n!}\langle\,:I^{n}\,{\rm e}^{-I}:\,\rangle,\;\;
I=\int_{0}^{t}dt'\,a^{{\rm out}\dagger}(t')\cdot a^{\rm out}(t'),\nonumber\\
&&a^{\rm out}(t)=(2\pi)^{-1/2}\int_{0}^{\infty}d\omega\,{\rm e}^{-{\rm i}\omega
t}a^{\rm out}(\omega).\label{PIarelation}
\end{eqnarray}
(The colons denote normal ordering.) It is convenient to work with the
generating function $F(\xi)=\sum_{p}\kappa_{p}\xi^{p}/p!$ of the factorial
cumulants $\kappa_{p}$ \cite{kappanote},
\begin{equation}
F(\xi)=\ln\sum_{n=0}^{\infty}(1+\xi)^{n}P(n)=\ln\langle\,:{\rm e}^{\xi
I}:\,\rangle.\label{Fxidef}
\end{equation}
To evaluate $F(\xi)$ we substitute Eq.\ (\ref{aoutSain}) into Eq.\
(\ref{PIarelation}) and perform the Gaussian averages.

A simple expression results in the long-time regime,
\begin{equation}
F(\xi)=-t\int_{0}^{\infty}\!\frac{d\omega}{2\pi}\,
\ln\bigl\|\openone-(\openone-S\cdot S^{\dagger})\xi f\bigr\|,\label{Fxilong}
\end{equation}
where $\|\cdots\|$ indicates the determinant. Eq.\ (\ref{Fxilong}) is valid
when $\omega_{\rm c}t\gg 1$, with $\omega_{\rm c}$ the frequency interval
within which $S\cdot S^{\dagger}$ does not vary appreciably. We have also found
a simple expression in the short-time regime,
\begin{equation}
F(\xi)=-\ln\bigl\|\openone-t\int_{0}^{\infty}\!
\frac{d\omega}{2\pi}\,(\openone-S\cdot S^{\dagger})\xi f\bigr\|,\label{Fxishort}
\end{equation}
valid when $\Omega_{\rm c}t\ll 1$, with $\Omega_{\rm c}$ the frequency range
over which $S\cdot S^{\dagger}$ differs appreciably from the unit matrix. (The
reciprocal of $\Omega_{\rm c}$ is the coherence time of the thermal emissions.)
The two equations (\ref{Fxilong}) and (\ref{Fxishort}) are the key results of
this paper. They reduce the quantum optical problem of the photon statistics to
a computation of the scattering matrix of the classical wave equation. That is
a major simplification, because the statistical properties of the scattering
matrix of a random medium are known from random-matrix theory
\cite{Bee97,Guh98}.

The long-time limit (\ref{Fxilong}) is particularly simple, as it depends only
on the set of eigenvalues $\sigma_{1},\sigma_{2},\ldots\sigma_{N}$ of $S\cdot
S^{\dagger}$. We call the $\sigma_{n}$'s ``scattering strengths''. An
additional simplification of the long-time regime is that one can do a
frequency-resolved measurement, counting only photons within a narrow frequency
interval $\delta\omega$ (with $\omega_{\rm c}\gg\delta\omega\gg 1/t$). The
factorial cumulants are then given by
\begin{equation}
\kappa_{p}=(p-1)!\,\nu
f^{p}N^{-1}\sum_{n=1}^{N}(1-\sigma_{n})^{p},\label{kapparesult}
\end{equation}
where $\nu=Nt\delta\omega/2\pi$ was defined in the introduction. For comparison
with black-body radiation we parameterize the variance in terms of the
effective number $\nu_{\rm eff}$ of degrees of freedom \cite{Man95},
\begin{equation}
{\rm Var}\,n=\bar{n}(1+\bar{n}/\nu_{\rm eff}),\label{nueffdef}
\end{equation}
with $\nu_{\rm eff}=\nu$ for a black body. Eq.\ (\ref{kapparesult}) implies
\begin{equation}
\frac{\nu_{\rm
eff}}{\nu}=\frac{\bigl[\sum_{n}(1-\sigma_{n})\bigr]^{2}}{N\sum_{n}
(1-\sigma_{n})^{2}}\leq 1.\label{nueffrho}
\end{equation}
We conclude that the super-Poissonian noise of a random medium corresponds to a
black body with a {\em reduced\/} number of degrees of freedom. Note that the
reduction occurs only for $N>1$.

We now turn to applications of our general formulas to specific random
media.  We concentrate on the long-time, frequency-resolved regime
with $N\gg 1$, leaving the short-time and single-mode regimes, and the
case of broad-band detection, for future publication \cite{Bee98}. An
ensemble of random media has a certain scattering-strength density
$\rho(\sigma)$. For $N\gg 1$ sample-to-sample fluctuations are small,
so the ensemble average is representative for a single system. We may
therefore replace $\sum_{n}$ by $\int d\sigma\,\rho(\sigma)$ in Eqs.\
(\ref{kapparesult}) and (\ref{nueffrho}).

\begin{figure}
\hspace*{\fill}
\psfig{figure=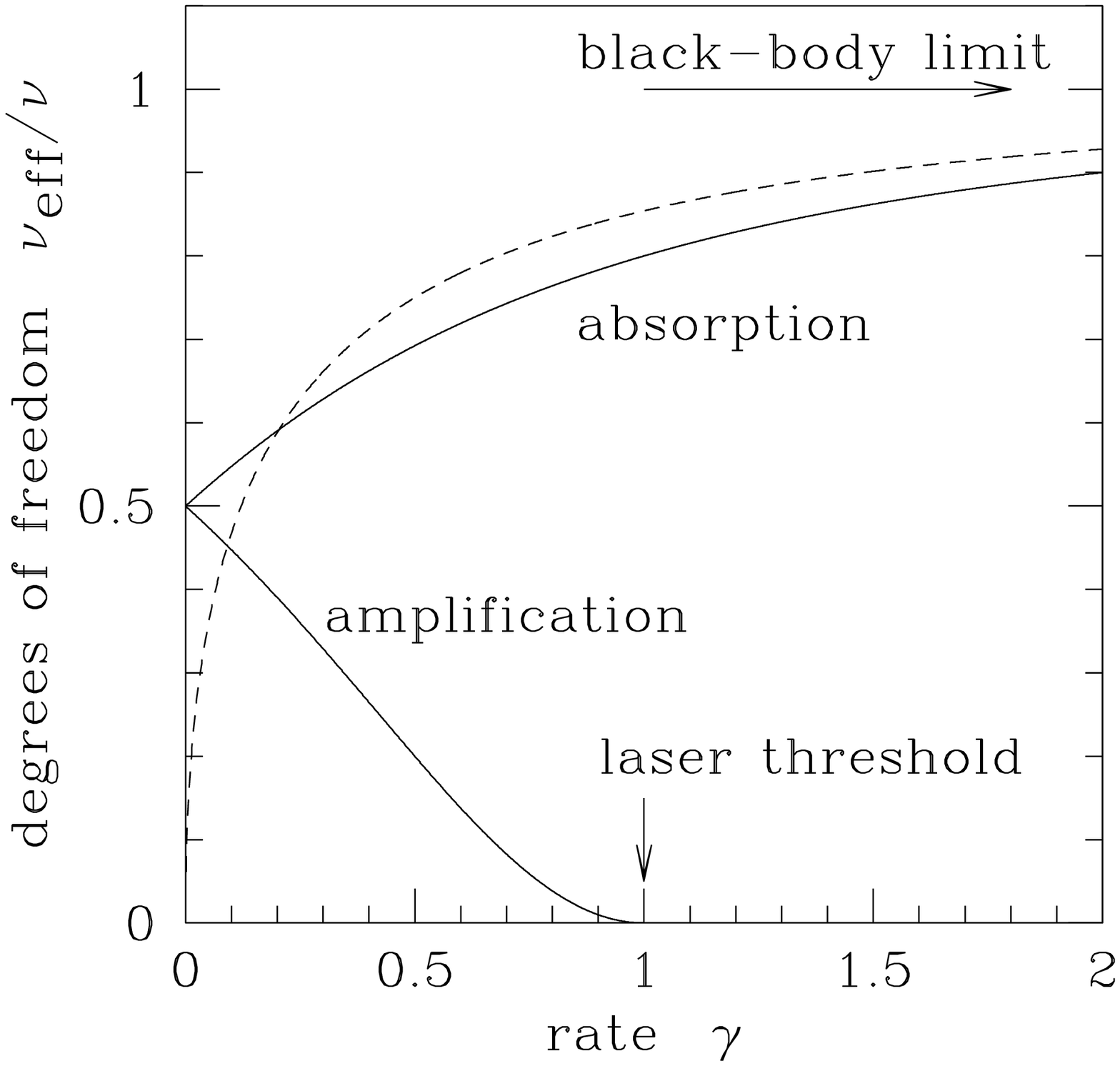,width=8cm}
\hspace*{\fill}\\
\refstepcounter{figure}
\label{nueff}
Fig.\ \ref{nueff} ---
Effective number of degrees of freedom as a function of normalized absorption
or amplification rate. The dashed curve is for the disordered slab, the solid
curves are for the chaotic cavity. The amplifying slab would be above the laser
threshold for any $\gamma$, so we only plot the case of absorption. For the
cavity both the cases of absorption and amplification are shown. The black-body
limit for absorbing systems and the laser threshold for amplifying systems are
indicated by arrows.
\end{figure}

As a first example we compute the thermal radiation from a disordered absorbing
slab. The slab is sufficiently thick that there is no transmission through it,
representing a semi-infinite random medium. We define the normalized absorption
rate \cite{coeff} $\gamma=\case{16}{3}\tau_{\rm s}/\tau_{\rm a}$. The
scattering-strength density $\rho(\sigma)$ in the regime $\gamma N^{2}\gg 1$ is
known \cite{Bee96,Bru96}. It is non-zero in the interval
$0<\sigma<(1+\case{1}{4}\gamma)^{-1}$, where it equals
\begin{equation}
\rho(\sigma)=(N/\pi)\sqrt{\gamma}(1-\sigma)^{-2}
(\sigma^{-1}-1-\case{1}{4}\gamma)^{1/2}.\label{rhoslab}
\end{equation}
This leads to the effective number of degrees of freedom
\begin{equation}
\nu_{\rm eff}/\nu=4[(1+4/\gamma)^{1/4}+(1+4/\gamma)^{-1/4}]^{-2},
\label{nueffslab}
\end{equation}
plotted in Fig.\ \ref{nueff}, with a mean photo\-count of
\begin{equation}
\bar{n}=\case{1}{2}\nu f\gamma\bigl(\sqrt{1+4/\gamma}-1\bigr).\label{barnslab}
\end{equation}
For strong absorption, $\gamma\gg 1$, we recover the black-body result
$\nu_{\rm eff}=\nu$, as expected. For weak absorption, $\gamma\ll 1$, we find
$\nu_{\rm eff}=2\nu\sqrt{\gamma}$. In the weak-absorption regime we can compute
the entire distribution $P(n)$ analytically. The result
\begin{equation}
P(n)\propto(\bar{n}^{n}/{n!})(1+f)^{-n/2} K_{n-1/2}\bigl(\nu_{\rm
eff}\sqrt{1+f}\bigr),\label{Pnslab}
\end{equation}
with $\bar{n}=\nu f\sqrt{\gamma}$ and $K$ a Bessel function, is Glauber's
distribution  \cite{Gla63} with a reduced number of degrees of freedom.

Our second example is an optical cavity connected to a photo\-detector via an
$N$-mode wave\-guide. The cavity modes near frequency $\omega$ are broadened
over a frequency range $N\Delta\omega$ much greater than their spacing
$\Delta\omega$ if $N\gg 1$. The cavity should have an irregular shape, or it
should contain random scatterers --- to ensure chaotic scattering of the
radiation. For this system we define the normalized absorption rate as
$\gamma=\tau_{\rm dwell}/\tau_{\rm a}$, where $\tau_{\rm dwell}\equiv
2\pi/N\Delta\omega\simeq 1/\omega_{\rm c}$ is the mean dwell time of a photon
in the cavity without absorption. The scattering-strength density for $N\gg 1$
follows from the general formulas of Ref.\ \cite{Bro96}. The result has a
simple form in the limit $\gamma\ll 1$ of weak absorption,
\begin{equation}
\rho(\sigma)=(N/2\pi)(1-\sigma)^{-2}(\sigma-\sigma_{-})^{1/2}
(\sigma_{+}-\sigma)^{1/2},\label{rhocavity}
\end{equation}
for $\sigma_{-}<\sigma<\sigma_{+}$ with $\sigma_{\pm}=1-3\gamma\pm
2\gamma\sqrt{2}$. In the opposite limit $\gamma\gg 1$ of strong absorption,
$\rho(\sigma)$ is given by the same Eq.\ (\ref{rhoslab}) as for the disordered
slab. We find the effective number of degrees of freedom
\begin{equation}
\nu_{\rm eff}/\nu=(1+\gamma)^{2}(\gamma^{2}+2\gamma+2)^{-1},
\label{nueffcavity}
\end{equation}
plotted also in Fig.\ \ref{nueff}, with a mean photo\-count of
\begin{equation}
\bar{n}=\nu f\gamma(1+\gamma)^{-1}.\label{barncavity}
\end{equation}
Again, $\nu_{\rm eff}=\nu$ for $\gamma\gg 1$. For $\gamma\ll 1$ we now find
$\nu_{\rm eff}=\case{1}{2}\nu$. It is remarkable that the ratio $\nu_{\rm
eff}/\nu$ for the chaotic cavity remains finite no matter how weak the
absorption, while this ratio goes to zero when $\gamma\rightarrow 0$ in the
case of the disordered slab.

These two examples concern thermal emission from absorbing systems. As we
discussed, our general formulas can also be applied to amplified spontaneous
emission, by evaluating the Bose-Einstein function (\ref{BEfunction}) at a
negative temperature. Complete population inversion corresponds to $f=-1$. A
duality relation \cite{Paa96} between absorbing and amplifying systems greatly
simplifies the calculation. The dielectric constants $\varepsilon'\pm{\rm
i}\varepsilon''$ of dual systems are each others complex conjugates, so dual
systems have the same value of $\tau_{\rm a}$ and $\gamma$. Their scattering
matrices are related by $S_{+}^{\dagger}=S_{-}^{-1}$, hence the scattering
strengths $\sigma_{1},\sigma_{2},\ldots\sigma_{N}$ of an amplifying system are
the reciprocal of those of the dual absorbing system.

We need to stay below the laser threshold, in order to be in the regime of
linear amplification. The semi-infinite medium is above the laser threshold no
matter how weak the amplification \cite{Bee96}, but the cavity is below
threshold as long as $\gamma<1$. We find that $\bar{n}$ and $\nu_{\rm eff}/\nu$
are given by Eqs.\ (\ref{nueffcavity}) and (\ref{barncavity}) upon substitution
of $\gamma$ by $-\gamma$. In Fig.\ \ref{nueff} we compare $\nu_{\rm eff}/\nu$
for amplifying and absorbing cavities. In the limit $\gamma\rightarrow 0$ the
two results coincide, but the $\gamma$-dependence is strikingly different:
While $\nu_{\rm eff}/\nu$ increases with $\gamma$ in the case of absorption, it
decreases in the case of amplification --- vanishing at the laser threshold. Of
course, close to the laser threshold [when $\gamma\gtrsim 1-(\Omega_{\rm
c}\tau_{\rm dwell})^{-1/2}$] the approximation of a linear amplifier breaks
down.

In summary, we have derived a relation between the photo\-count distribution
$P(n)$, in the long-time limit, and the eigenvalues
$\sigma_{1},\sigma_{2},\ldots\sigma_{N}$ of the scattering-matrix product
$S\cdot S^{\dagger}$. The super-Poissonian noise ${\rm
Var}\,n=\bar{n}(1+\bar{n}/\nu_{\rm eff})$ is that of a black body with a
reduced number $\nu_{\rm eff}$ of degrees of freedom. We have computed
$\nu_{\rm eff}$ for several types of random media, in the large-$N$ regime,
using results from random-matrix theory. In a weakly absorbing or amplifying
chaotic cavity, the ratio $\nu_{\rm eff}/\nu$ is a universal factor of $1/2$
--- independent of microscopic parameters. In a disordered slab, $\nu_{\rm
eff}/\nu$ vanishes $\propto 1/\sqrt{\tau_{\rm a}}$ for small absorption rates
$1/\tau_{\rm a}$. We have found that $\nu_{\rm eff}/\nu$ vanishes also on
approaching the laser threshold in an amplifying chaotic cavity.

The reduction of $\nu_{\rm eff}$ amounts to an excess noise of amplified
spontaneous emission. Its origin is the presence of a large number $N$ of
overlapping cavity modes, and a broad distribution $\rho(\sigma)$ of the
corresponding scattering strengths. Overlap of cavity modes is avoided in the
usual laser geometry, but it is generic in a random laser. This fundamental
difference was pointed out thirty years ago by Letokhov \cite{Let67}, in the
paper that pioneered the notion of a ``stochastic resonator''. Letokhov
concludes his paper by surmising that the statistical properties of spontaneous
emission would be distinctly different from the usual case. The reduction of
the number of degrees of freedom predicted here forms an experimentally
accessible signature of this difference.

I have benefitted from discussions with P. W. Brouwer, M. P. van Exter, and J.
P. Woerdman. This work was supported by the Dutch Science Foundation NWO/FOM.

\ecols

\begin{references}
\bibitem{alpha} The formulas in the text are for the case of detection
efficiency $\alpha=1$ photo\-electron per photon. If $\alpha<1$, one should
replace $f$ by $\alpha f$.
\bibitem{kappanote} The factorial cumulants $\kappa_{p}$ are the cumulants of
the factorial moments $\overline{n(n-1)\cdots(n-p+1)}$. For example,
$\kappa_{2}=\overline{n(n-1)}-\bar{n}^{2}$.
\bibitem{coeff} The coefficient in the definition of $\gamma$ is chosen to
facilitate the comparison between slab and cavity.
\end{references}
\end{document}